\documentclass[doublecol]{epl2}

\title{Field dependence of the quantum ground state in the Shastry-Sutherland system
SrCu$_2$(BO$_3$)$_2$}
\shorttitle{Field dependence of the quantum ground state in SrCu$_2$(BO$_3$)$_2$} 

\author{
F. Levy\inst{1} \and I. Sheikin\inst{1} \and C. Berthier\inst{1} \and M.
Horvati\'c\inst{1} \and M. Takigawa\inst{2} \and H. Kageyama\inst{3} \and T.
Waki\inst{2} \and Y. Ueda\inst{2} } \shortauthor{F.Levy \etal}

\institute{
 \inst{1} Grenoble High Magnetic Field Laboratory
 (GHMFL) - CNRS, BP 166, 38042 Grenoble Cedex 09, France \\
 \inst{2}
 Institute for Solid State Physics, University of Tokyo,
 Kashiwanoha, Kashiwa, 277-8581, Japan \\
\inst{3}Department of Chemistry, Graduate School of Science, Kyoto
 University, Kyoto 606-8502, Japan
}

 {Quantized spin models}  {Magnetic
phase boundaries}  {Magnetocaloric effect, magnetic
cooling}

\abstract{We present magnetic torque measurements on the Shastry-Sutherland
quantum spin system SrCu$_2$(BO$_3$)$_2$ in fields up to 31~T and temperatures
down to 50~mK. A new quantum phase is observed in a 1~T field range above the
1/8 plateau, in agreement with recent NMR results. Since the presence of the DM
coupling precludes the existence of a true Bose-Einstein condensation and the
formation of a supersolid phase in SrCu$_2$(BO$_3$)$_2$, the exact nature of
the new phase in the vicinity of the plateau remains to be explained.
Comparison between magnetization and torque data reveals a huge contribution of
the Dzyaloshinskii-Moriya interaction to the torque response. Finally, our
measurements demonstrate the existence of a supercooling due to adiabatic
magnetocaloric effects in pulsed field experiments.}

\begin{document}

\maketitle

Quantum antiferromagnetic spin systems with singlet ground states exhibit a
variety of magnetic field induced quantum phase transitions.   Crystalline
arrays of $S=1/2$ spin dimers, for instance, can present two contrasting
behaviors \cite{Rice_02}. When the magnetic field exceeds a critical value at
which the lowest energy levels cross each other, the triplet excitations, which
can be treated as hard core bosons on a lattice, typically undergo a
Bose-Einstein condensation (BEC) \cite{Giamarchi_99,Nikuni_00,Jaime_04}.
Another possibility, however, is the occurrence of magnetization plateaus at
fractional values of the saturated magnetization. Such plateaus correspond to
the formation of a superlattice of triplets (``magnetic crystal'') and may
occur when the kinetic energy of the triplets is strongly reduced by
frustration, so that the repulsive interactions become dominant. The best known
example for the formation of such plateaus is SrCu$_{2}$(BO$_{3}$)$_{2}$ with
its two-dimensional network of orthogonal dimers of $S$ = 1/2 Cu$^{2+}$ ions
\cite{Kageyama_99}. This material shows an excitation gap $\Delta_{0}=35$~K and
plateaus at 1/8, 1/4, and 1/3 of the saturated magnetization
\cite{Onizuka_00,Miyahara_03}. A magnetic superlattice at the 1/8-plateau has
actually been observed in NMR experiments \cite{Kodama_02}. It has been argued
on theoretical basis that some analog of a supersolid phase
\cite{Momoi_00,Ng_06,Laflorencie_07,Schmidt_07}, consisting of the
superposition of the magnetic crystal and a Bose-Einstein condensate of the
interstitial triplets, could occur in the vicinity of the plateau phases. The
presence of such an exotic phase, the magnetic analog of the highly debated
supersolid phase in $^4$He, is however precluded in SrCu$_{2}$(BO$_{3}$)$_{2}$
because of the presence of an intradimer Dzyaloshinskii-Moriya (DM) interaction
\cite{Zorko_04,Kodama_05} which breaks the U(1) symmetry. However, NMR
measurements have recently revealed the existence of new magnetic phases above
the 1/8 plateau. This prompted us to reexamine the field temperature ($H$-$T$)
phase diagram of SrCu$_{2}$(BO$_{3}$)$_{2}$ up to 31~T using torque
measurements. Our experiments indeed confirm the existence of a new phase
adjacent to the 1/8 plateau, in which the magnetization is only slowly
increasing. In addition, we report for the first time the field dependence of
the pure longitudinal magnetization up to 31~T, measured at the temperature of
60~mK.  The results strongly differ from those obtained by torque measurements,
as expected in presence of DM interaction within the dimers \cite{Miyahara_07}.
In particular, the magnetization jump before the 1/8 plateau is much larger
than previously reported \cite{Onizuka_00, Jorge_05}, in excellent agreement
with NMR data.

The torque measurements were performed at the Grenoble High Magnetic Field
Laboratory in a 20 MW resistive magnet equipped with a dilution refrigerator.
The sample ( $\sim$~$1 \times 1 \times 0.5$~mm$^{3}$ size) was mounted on a
25~$\mu$m thick CuBe cantilever with its $c$-axis perpendicular to the surface.
In situ rotation allowed us to obtain an angle $\theta$ between the $c$-axis
and the applied magnetic field of about 0.4$^\circ$, a configuration which has
been used for most of the experiments. Torque measurements have been performed
at various constant temperatures while sweeping field up (at a rate of
100~Gauss/s) and down (at a rate of 200~Gauss/s).

\begin{figure}[t]
\begin{center}
\resizebox{6cm}{6cm}{\includegraphics{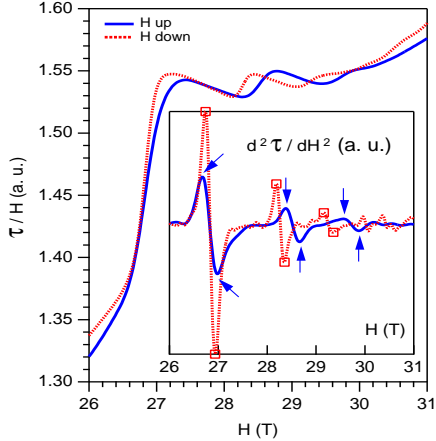}} \caption{\textbf{Torque
divided by field versus field.} The blue continuous line corresponds to a field
sweep up at 55 $\pm$ 10~mK. The red dashed line corresponds to a field sweep
down at 90 $\pm$ 10~mK. The inset shows the corresponding second derivatives.
Six extrema (three transitions) are found in each sweep and pointed out by
arrow for field sweep up, and by squares for field sweep down. } \label{fig1}
\end{center}
\end{figure}
When the sample is placed in a homogeneous field, the torque is proportional to
the component of the magnetization perpendicular to the applied field and to
the rotation axis of the cantilever. However, if the sample is placed in a
field gradient, one additionally obtains an access to the magnetization
parallel to the applied field, provided the torque component becomes negligible
with respect to the force $\mathbf{F}= -\mathbf{M}
\mu_0\vec{\bigtriangledown}H$. For this we moved the sample by 1~cm above the
magnet center. From here on, we will refer to such measurements as ``true''
magnetization measurements.

Figure~\ref{fig1} shows the results obtained at the lowest temperature. Three
anomalies are clearly visible in the second derivative of the torque divided by
field $\tau/H$. The first two  correspond to the boundaries of the 1/8 plateau.
Above the plateau, there is a second phase which ends up around 29.5~T. In both
phases, $\tau/H$ is slightly decreasing with increasing the field. Comparing
data acquired in ramping up and down the magnetic field, one observes  no
hysteresis for the lower boundary of the plateau. This absence of hysteresis as
a function of $H$ has already been observed in NMR experiments
\cite{Kodama_02}, in spite of the fact that the transition is of the first
order. This contrasts with the two other transitions, which exhibit a rather
strong hysteresis.

\begin{figure}[t]
\begin{center}
\resizebox{6cm}{6cm}{\includegraphics{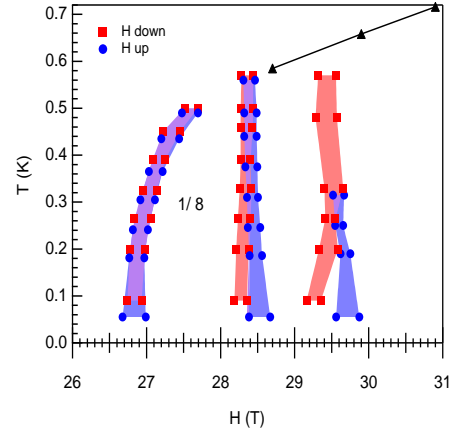}}
\caption{\textbf{Temperature-field phase diagram.}  Position of the  extrema of
the second derivative of the torque measurements made at various constant
temperature (see fig.~\ref{fig1}). Three transition lines are clearly
appearing. The blue circles correspond to field sweep up and the red squares to
field sweep down. The NMR points, black triangles, delimit the boundary of the
magnetic ordering. } \label{fig2}
\end{center}
\end{figure}

The position of the peaks found in the second derivative of the torque (see
fig.~\ref{fig1}) are reported in fig.~\ref{fig2} to establish the
field-temperature phase diagram.  The transition temperatures delimiting  the
phase boundary between the paramagnetic  and the ordered magnetic states (black
triangles) are taken from NMR results \cite{Takigawa_07}. The field values
corresponding to the lower and the upper boundary of the plateau are in
excellent agreement with the NMR data: at very low temperature, the coexistence
between the uniform paramagnetic phase and the triplet superlattice
 starts at 26.6~T and the uniform
phase  disappears at 27~T \cite{Kodama_02}. The transition from the 1/8 plateau
into the adjacent phase above was found to start at 28.3~T at 0.31~K both by
NMR \cite{Takigawa_07} and torque measurements.

\begin{figure}[t]
\begin{center}
\resizebox{6cm}{6cm}{\includegraphics{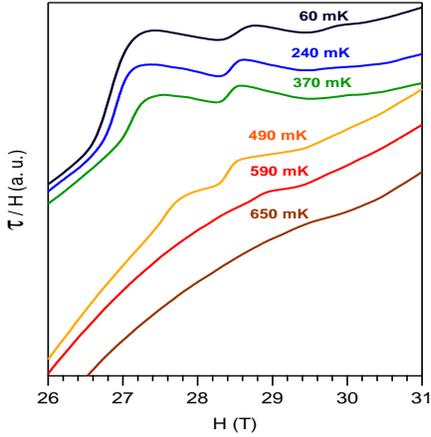}}
\caption{\textbf{Torque divided by field versus field at different
temperatures.} The measurements were made for field sweep up. The
curves have been arbitrary shifted
 for clarity.
} \label{fig3}
\end{center}
\end{figure}

\begin{figure}[t]
\begin{center}
\resizebox{6cm}{6cm}{\includegraphics{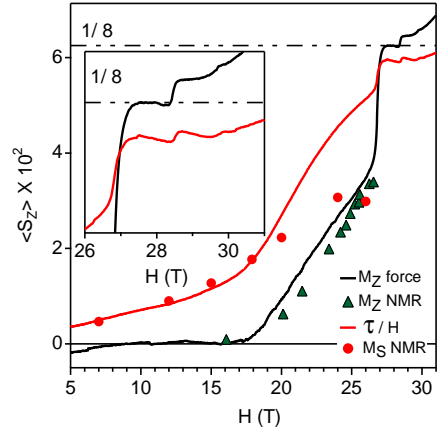}} \caption{\textbf{Comparison of
torque and magnetization.} Magnetization was measured both by NMR technique,
green triangles, and by torque in a field gradient, continuous black line.
Those measurements are compared with  torque divided by field, the continuous
red line. The solid red circles correspond to the transverse staggered
magnetization measured by $^{11}$B NMR.} \label{fig4}
\end{center}
\end{figure}

We now consider the temperature dependence of $\tau/H$ shown in
fig.~\ref{fig3}. There are two remarkable features in these data. First, one
can see that the signature of the 1/8 plateau has fully disappeared at 590~mK,
while there is still some reminiscence of the adjacent phase at this
temperature. This again is in excellent agreement with previous NMR data
\cite{Kodama_02}, but contrasts with earlier measurements made in pulsed
magnetic field performed at 1.4~K \cite{Onizuka_00,Jorge_05}, in which the
signature of the plateau is still visible. Since the gap between the lowest
triplet excitations branch and the singlet state decreases toward a very small
value \cite{Kodama_05, Nojiri_03} as the magnetic field is approaching the
value corresponding  to the entry of the plateau, the ``high temperature''
observation of the plateau in pulsed magnetic field can be interpreted in terms
of adiabatic (isentropic) cooling of the spin system
\cite{Honecker_06,Honecker_unpublished}.
As long as  only the lowest triplet branch and the singlet state are necessary
to describe the system, this is an analog of the cooling of paramagnetic salts
by adiabatic demagnetization. The second remarkable feature in fig.~\ref{fig3}
is that between 370~mK and 490~mK the slope of $\tau/H$ changes from negative
to positive before the complete melting of the magnetic superlattice. This
corresponds to the regime in which the triplet superlattice and the
paramagnetic phase coexist, which was found by NMR to extend between 360~mK and
520~mK at 27.6~T \cite{Takigawa_04}. Such a change of sign, related to the
coexistence of the superlattice and the uniform phase, can only be explained if
the contribution to $\tau/H$ of the superlattice has a negative slope with
increasing $H$, while that of the uniform paramagnetic phase has a positive
one. This is rather surprising, since within the plateau one would expect the
torque signal to remain constant. In order to elucidate this issue, we have
performed a ``true magnetization'' measurement in which the bending of the
cantilever is now dominated by the force $F_{z} = - M_{z} dB_{z}/dz$. The
results are shown in fig.~\ref{fig4}, together with a torque measurement
recorded at the same temperature.
In addition, solid triangles show the amplitude of $\langle S_{z}\rangle$ as
determined by NMR. The field dependence of the longitudinal magnetization
$M_{z}$ actually strongly differs from the results obtained by torque. $M_{z}$
is \emph{flat} within the 1/8 plateau, as expected, and within the adjacent
phase between 28.4 and $\simeq$29.5~T, it is nearly flat with only a small
increase as approaching the upper boundary. The magnetization data are in
excellent agreement with the NMR results. In particular, both techniques reveal
a large jump of the magnetization just before the 1/8 plateau, in contrast to
all previously reported data \cite{Onizuka_00,Jorge_05}.

We also remark that recently it has been proposed that the 1/8 plateau could be
preceded by a 1/9 plateau, and followed by a 1/7 plateau \cite{Sebastian_07}.
Indeed, the authors of Ref.~\cite{Sebastian_07} report three transition lines
around 27.1, 29 and 30.3~T.  In turns out that by reducing these three fields
values by 2~\% we recover the phase boundaries values reported in this letter.
These latter values agree with those determined by NMR, a technique which
inherently always provides a precise determination of the magnetic field.
Therefore, the values reported in Ref.~\cite{Sebastian_07} appear to result
from an incorrect field scale. Furthermore, looking carefully at the pulsed
field measurements of Ref~\cite{Onizuka_00}, the magnetization values measured
for the 1/4 and 1/3 plateaus clearly indicate that the phase extending between
26.7 and 28.3~T can only correspond to the 1/8 plateau. As far as the adjacent
phase is concerned, our measurements of the longitudinal magnetization
demonstrate that it cannot correspond to a 1/7 plateau, since the increase of
$M/M_{sat}$ is too small.

\begin{figure}[t]
\begin{center}
\resizebox{5 cm}{!}{\includegraphics{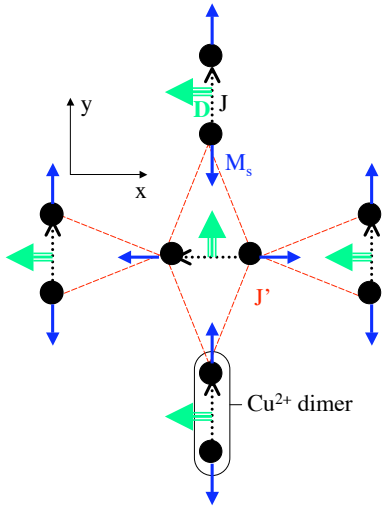}} \caption{\textbf{Schematic
structure of the Cu$^{2+}$ dimers.} The figure sketches the orthogonal network
of Cu$^{2+}$ dimers. Black circles represent the Cu$^{2+}$ ions.  Intradimer
interactions are represented by  dotted black lines, and interdimer interaction
J' is by dashed red ones. Thick green arrows indicate the direction of
$\mathbf{D}$ vectors for the intradimer Dzyaloshinskii-Moriya interaction
$\mathbf{D} \cdot \mathbf{S}_i \times \mathbf{S}_j$, where the bond
``direction'' $i \rightarrow j$ is shown by black dotted arrows. The solid blue
arrows represent the staggered magnetization M$_s$ induced by a field applied
along the c-axis (z-axis), within the field range below the first magnetization
plateau. Interdimer DM interactions, which are less effective to generate a
transverse magnetization, are not shown here.} \label{Cu_plane}
\end{center}
\end{figure}

SrCu$_{2}$(BO$_{3}$)$_{2}$ crystallizes in a tetragonal structure with
alternative layers of Sr and Cu(BO$_{3}$) planes along the $c$-axis. At low
temperature, due to the buckling of the BO$_3$-Cu-O-Cu-BO$_3$ bonding, the $ab$
plane is no longer a mirror plane \cite{Sparta_01}, allowing the existence of
an in-plane intradimer DM interaction as shown on fig.~\ref{Cu_plane}, as well
as a staggered $g$-tensor. Interdimer DM interactions are also present
\cite{Cepas_01}, but their role  is mainly  to partially restore some kinetic
energy to the triplets. The strong difference between the longitudinal
magnetization $M_{z}(H)$ and the torque signal as shown in fig.~\ref{fig4} is
indeed the signature of the intradimer DM interaction and the staggered $g$
tensor. The presence of the intradimer DM interaction has been shown to
generate a transverse staggered magnetization, observed by $^{11}$B NMR  and
computed by exact diagonalization \cite{Kodama_05}. While this transverse
staggered magnetization has no effect on the torque,  it has been shown
recently that there is an additional \emph{uniform} transverse component
generated by the DM interaction, which is smaller by an order of $D/J$
\cite{Miyahara_07}. In the low field limit and for an isolated dimer, this
component for each dimer has the symmetry of $\mathbf{D}\times \mathbf{D}\times
\mathbf{H}$. Within the low field approximation, the torque per spin dimer can
be expressed as:
\begin{eqnarray}
\nonumber
 \tau = (\chi_{ab}-\chi_{c})\sin\theta\cos\theta
H^{2} -g\mu_{\rm{B}}D^{2}/4J^{3}\sin\theta\cos\theta H^{2},
\end{eqnarray}
where $\theta$ is the angle between the $c$-axis and the applied magnetic
field, and $\mathbf{\chi}$ is the part of the susceptibility which only depends
on the symmetric part of the $g$-tensor.

The first term is the standard contribution which is proportional to the
longitudinal magnetization for small values of $\theta$, while the second one
results from the DM interaction. Both terms have the same angular dependence,
and vanish for $\theta = 0$ ($H \| c$). In fig.~\ref{fig4} the field dependence
of the torque signal is compared to that of the staggered magnetization
determined from NMR measurements \cite{Kodama_05}. In the low field limit, in
which the torque signal is only (as long as $M_z$ = 0) or mainly due to the DM
interaction,  their variation is quite similar, which is in agreement with the
theory predicting that both quantities vary linearly with $H$. When $M_z \neq $
0 the torque becomes the sum of two contributions and a direct comparison is no
longer possible. Exact diagonalization calculations are required to determined
the full field dependence of the ``uniform'' transverse magnetization in the
Shastry-Sutherland geometry.

\begin{figure}[t]
\begin{center}
\resizebox{6cm}{6cm}{\includegraphics{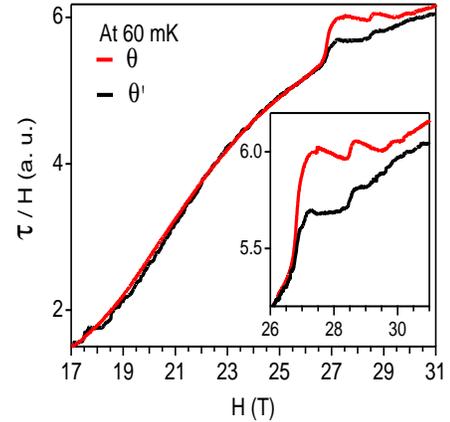}}
\caption{\textbf{Dzyaloshinski-Moriya contribution within the 1/8 plateau and
its adjacent phase.} Torque data have been recorded at $\theta$ = +0.4$^\circ$
and at $-|\theta'|$. The latter data have been rescaled to give the same signal
at 17 and 26~T. Two signals strongly differ within the plateau and its adjacent
phase, which is attributed to a symmetry breaking. } \label{fig6}
\end{center}
\end{figure}

We now consider the $\tau /H$ variation within the plateau and its adjacent
phase and try to understand the origin of the negative slope observed within
the 1/8 plateau. Figure~\ref{fig6} shows torque measurements recorded at
$\theta$ and at $-|\theta'|$. Since, as expected, the corresponding raw data
have different sign, those corresponding to $-|\theta'|$ have been renormalized
in order to give the same values at 17 and 26~T. One can see that the field
variation of both signals are identical in the uniform phase, as expected if
they only differ by a factor $\sin\theta\cos\theta$. However, they strongly
differ within the 1/8 plateau and its adjacent phase, in which an extra
contribution is observed. This is expected, at least for the 1/8 plateau, for
the following reason. Within the uniform phase, the DM interaction is conserved
by the three symmetry operations of the crystallographic structure: the mirror
plane $zx$, the mirror plane $yz$ and a C2 rotation at the intersection of
these mirror planes. However, within the 1/8 plateau, the structure determined
for the magnetic superlattice \cite{Kodama_02} has only one symmetry left,
which is a C2 rotation around the middle of the most polarized dimer. We thus
expect that the angular dependence of the torque signal becomes different, and
starts to depend also on the angle between the projection of $H$ on the $ab$
plane and the crystallographic axes.
Recently, the investigation of frustrated ladders with DM interactions in a
magnetic field \cite{Penc_07} has been extended to the situation where the
field is neither parallel nor perpendicular to the $\mathbf{D}$ vector
\cite{Manmana_07}. It has been shown that the torque induced by the DM
interaction develops peaks upon entering and leaving the 1/2 magnetization
plateau. While the torque produced by the misalignment of the field with a
principal axis of the $g$-tensor increases monotonously with the field, the
torque induced by the DM interaction is non monotonous inside the plateau, in
qualitative agreement with the present observation. Whether this anomalous
contribution disappears or not above 29.5~T, where we know from NMR that the
``magnetic crystal'' persists \cite{Takigawa_07}, is not clear at the moment,
and would require new measurements.

What is the nature of the phase adjacent to the 1/8 plateau?
Recent NMR experiments have shown that the magnetic superlattice,
analogous to a  magnetic crystal, does not melt when additional
triplets are introduced. One can then immediately suspect that
this new phase is the analog of a supersolid phase, in which the
additional triplets would undergo a Bose-Einstein condensation.
However, the presence of the intradimer DM interaction and the
resulting staggered magnetization break the U(1) symmetry around
the applied magnetic field and thus remove the continuous symmetry
of a supersolid phase. So, some more sophisticated theoretical
description of the exact nature of this phase has to be provided
in the future.

In conclusion, we have determined the phase diagram of the Shastry-Sutherland
quantum antiferromagnet SrCu$_{2}$(BO$_{3}$)$_{2}$ in the $H$-$T$ plane up to
31~T, using both magnetic torque and ``pure longitudinal'' magnetization
measurements. We show that the torque measurements allow the detection of the
phase transitions between successive quantum ground states, but cannot give
access to the true variation of the longitudinal magnetization $M_{z}$.  This
is due to the existence of an intradimer Dzyaloshinskii-Moriya interaction,
which generates an additional uniform transverse magnetization providing a
strong contribution to the torque. In the low field limit this contribution
scales linearly with the transverse staggered magnetization measured by NMR.
The phase boundaries of the 1/8 magnetization plateau are found to be in
agreement with NMR data: at 60~mK, the coexistence between the uniform
paramagnetic phase and the 1/8 plateau extends from 26.6 to 27~T, and the
plateau ends through a first order transition starting at 28.3~T. The
temperature  corresponding to the complete melting of the spin superlattice in
the 1/8 plateau is at most 570~mK. This demonstrates that the observation of
the 1/8 plateau at 1.4~K in pulsed field measurements is due to isentropic
adiabatic cooling of the spin system. The most important finding of this study
is the evidence of a new phase adjacent to the 1/8 plateau and extending up to
29.3~T. The magnetization within this phase is nearly field independent and
only increases when approaching its upper boundary. However, its value does not
correspond to any simple rational value of the saturation magnetization. Recent
NMR measurements show that the ``magnetic crystal'' does not melt in that
phase. However, since the Dzyaloshinskii-Moriya interaction removes the
continuous rotation symmetry in SrCu$_{2}$(BO$_{3}$)$_{2}$, this phase has to
be more complex than a simple analog of  a supersolid, which would correspond
to the coexistence of the 1/8 plateau ``magnetic crystal'' and a Bose-Einstein
condensate of the interstitial triplets. This clearly shows that our
understanding of the physics of interacting hard core bosons on a lattice has
to be improved, and that further theoretical and experimental investigation are
necessary to clarify the evolution of the quantum ground states between the 1/8
and the 1/4 plateaus in this model compound.

\acknowledgments We acknowledge F. Mila for enlightening
discussions. This work was supported by the the French ANR grant
06-BLAN-0111 and Grant-in-Aids for Scientific Research (Nos.
16076204 and 19052004) from the MEXT Japan.

\end{document}